\documentclass[12pt]{article}
\usepackage{amssymb}
\begin{document}
\title{\bf A new class of scale free solutions to linear ordinary 
differential equations and the universality of the Golden Mean 
${\frac{\sqrt{5}-1}{2}}$}

\author{
Dhurjati Prasad Datta\\
Department of Mathematics \\
North Eastern Regional Institute of
Science and Technology\\
Itanagar-791109, Arunachal Pradesh, India} 
\date{}
\maketitle
\begin{center}email:dp$_-$datta@yahoo.com\end{center}

\baselineskip = 20pt

\begin{abstract}

A new class of finitely differentiable scale free  solutions to the
simplest class of ordinary 
differential equations is presented. Consequently, the real number set
gets replaced 
by an extended physical set, each element of which is endowed with an
equivalence 
class of infinitesimally separated neighbours in the form of random
fluctuations. We show 
how a sense of time and evolution is intrinsically defined by the infinite
continued fraction 
of the golden mean irrational number ${\frac{\sqrt{5}-1}{2}}$, which plays
a key role in this 
extended SL(2,R) formalism of Calculus. Time may thereby undergo random
inversions 
generating well defined random scales, thus allowing a dynamical  
system to evolve self similarly over the set of multiple scales. The late
time 
stochastic fluctuations of a dynamical system enjoys the generic $1/f$
spectrum. A universal 
form of the related probability density is also derived. We prove that the
golden mean 
number is intrinsically random, letting all measurements in the physical
universe 
fundamentally uncertain. The present analysis offers an explanation of the
universal 
occurrence of the golden mean in diverse natural and biological processes.
\end{abstract}
\begin{center} PACS Numbers: 02.30.Hq; 05.45.+j; 47.53.+n 
\end{center}
\newpage
\section{Introduction} 

In Calculus,  a variable changes by ordinary shift operations 
(translations). Every point in the real axis is conceived as a
structureless point. In a 
dynamical problem, the time evolution of a system occurs over a finite
number of 
characteristic scales as inherited from the underlying differential
equation. The 
generation of fractal- like self similar structures over multiple
(dynamically generated) 
scales, either in the phase space or purely in the time sector, needs
explicit nonlinearity 
at the level of the governing equation.  
In the following, we present a new class of scale free stochastic
solutions of the simplest 
ordinary differential equation. A representative solution of the new
class, which is exact and 
finitely differentiable, defines an extension of the real number set,
endowing every number 
with a nontrivial neighbourhood of fluctuations. These fluctuations
experience an intrinsic 
universal evolution, which, in turn, generates a generic irreversible
sense of time 
defined by the infinite continued fraction representation of the `golden
mean' irrational number. 
Consequently, time may undergo random inversions, allowing a dynamical
system to 
explore stochastically, new scales, usually unavailable in the ordinary
dynamics. The 
corresponding evolution of the system would, therefore, enjoy a generic
late time stochastic  
fluctuations with  $1/f$ spectrum [1]. We also obtain an exact form of the
universal 
probability density, recently found [2] to occur naturally in a wide class
of scale free 
nonlinear processes in Nature.  These exact results provide a natural
explanation 
of the universal occurrence of the golden mean in diverse natural and
biological processes.

\par The relevance of the present work may also be seen in the light 
of the recent works [3,4] uncovering new relationships 
between time and the number theory. In ref [3], the evidence 
of 1/$f$ noise and scale free self similarity in the prime number 
distribution is pointed out. In ref [4], Planat reported 
the observation of discrete time jumps in beat frequencies 
in the context of a superheterodyne receiver. These random 
jumps are shown to generate 1/$f$ noise in the oscillator 
frequencies. Further, the origin of 1/$f$ spectrum is related 
to the arithmetical summatory functions such as the Mobius 
and the Mangoldt functions which arise in connection with 
the Reimann zeta function. Sometime back,  Robinson, on the other-hand, 
developed [5] a theory of infinitesimals on the basis of 
mathematical logic, thereby giving rise to a valid model 
of a nonstandard extension of the standard framework of 
analysis. As it turns out, the present extension of 
Calculus is a realization of the nonstandard analysis, 
in which infinitesimally small numbers are shown to have 
dynamical properties. We note also that our exact results 
seem to provide a correct mathematical framework for 
the theories of fractal space-time being developed by 
several authors [6,7,8] . A significant 
part of El Naschie's work [8], in particular, explores the significance of 
the golden mean in Cantorian fractal spacetime and other related 
branches of Physics and Mathematics. 
A detailed comparison 
of the present results with those of Refs.[6-8] would require separate 
investigations. Incidentally,  the possibility of fractal solutions 
to linear differential equations seems to have been pointed out first by 
Nottale [6].\footnote{Prof. El Naschie has pointed out that it is in-fact
G. N. Ord 
who first conceived the possibility of fractal solutions to  linear
differential equations.}

\par To indicate how the exact results, reported here, are obtained,
starting 
from a heuristic definition of time inversion, we present the exact
results in Sec.4. In the  
preceding two sections we indicate  how the exact class of solutions is
derived in Sec.3, using an  
approximate analysis, based on a local definition of time inversion,
explained in Sec.2. 
Applications of the results are discussed in Sec.5. We close the
presentation with some 
further remarks pointing out the future scope of  this extended formalism
of Calculus.

\section{Time inversion}
\par Let us consider the simplest linear dynamical system given by

\begin{equation}\label{simo}
{{{\rm d}x\over {\rm d}t}=(1+\kappa t)x}
\end{equation}

\noindent where $t$ denotes the dimensionless time ( we scale
$t$ suitably to adjust the dominant scale of evolution to $t\sim 1$) and 
$\kappa$ is a small, slowly varying (almost constant), parameter.  
In the ordinary Calculus, this equation has the `standard' solution
$x_s\propto \exp(t+
{\frac{1}{2}}\kappa t^2)$, with no (self-similar) fluctuations, 
unless the equation is explicitly nonlinear, for instance,  through
$\kappa 
=\kappa (t,x,\dot x,\ddot x,\ldots)$. Our intention is to show that
eq~(\ref{simo}) 
can accommodate a new class of nonlinear stochastic solutions, even for 
a constant $\kappa $, under a simple, but  general assumption that 
{\em time may change by inversions as well}.

Let us recall that  ordinarily a change in time, in the vicinity of a
given 
instant $t_0$ is indicated by a pure translation $t=t_0+\bar t\equiv
t_0+(t-t_0)$. 
In fact, the last equality is an identity (valid for all $t$). 
By an inversion, on the other hand, we mean the following.
Let $t_{\pm}$ denote times $t\lessapprox 1$ and $t\gtrapprox1$
respectively 
(from now on $t$  denotes the rescaled variable $t\rightarrow
{\frac{t}{t_0}}$). 
Then close to $t=1$, the inversion $t_{-}=1/(1+(t_{+}-1))$
leads to the {\em constraint} $1-t_{-}=t_{+}-1$. The parametric
representation
of inversely related times is obviously given by $t_{-}=1-\bar t$ and
$t_{+}=1+\bar t$, $0<\bar t<<1$ (so that the constraint reduces to an {\em 
identity}, valid close to $t=1$). With this reinterpretation, time
inversion
in the vicinity of $t=1$ assumes a form analogous  to a pure translation.
If translation is considered to be the most natural mode of time
increment, 
then there is no compelling reason of ignoring inversion
as yet another {\em natural} mode of doing this. Consequently, it seems 
reasonable to assume that time may change from $t_{-}$ to $t_{+}$ not only
by
ordinary translation over the period $t_{+}-t_{-}=2\bar t$, but also 
instantaneously by an inversion. Let us 
 clarify the physical content of the definition further through several
remarks.

1. Note that the above definition gives 
a {\em new nontrivial} solution $t_{-}=1/t_{+}$ to the constraint
$t_{-}+t_{+}=2$, in the {\em vicinity} of $t=1$ over the linear solution 
$t_{-}=2-t_{+}$, ordinarily thought to be the only possible solution. 
In view of this nonlinear possibility, the change (flow) of 
time could be visualised as an SL(2,R) group action, when the nontrivial
SL(2,R) action is realized only in a neighbourhood of a point, $t=1$, say. 
One may thus imagine that time flows, for example,  from 
$t=0$ to $t=t_{-}$ by translation, and then may switch over to $t_{+}$ by 
inversion $t_{+}=1/t_{-}$, for another period of linear flow etc. In the
next sections, 
we show how this second and subsequent periods of linear flows 
are actually realized over scales of the form $t_n=\nu^n
t,\,n=1,2,\ldots$, where 
$ \nu={\frac{\sqrt{5}-1}{2}}=0.618033939...$, is the  golden mean
irrational 
number.\footnote{The standard notation of the golden mean in the
literature is 
$\phi$. Our symbol $\nu$ in this work is motivated by the fact that the
golden mean  
appears here as a universal scaling factor of time. The symbol $\phi$ is
more 
often used to denote a quantum state function as well. In Sec.4-5, we,
however, denote the 
stochastic time-like character of the golden mean by $\phi$.}   

2. The definition of time inversion has an inbuilt uncertainty. The
moments $t_{\pm}$ 
are not well defined, except for the fact that these be close to $t=1$,
thus elevating
$\bar t$, and hence time itself,  to the status of a random variable. Note
that this 
sort of uncertainty (randomness) is not allowed in the framework of linear
Newtonian 
time. In the present SL(2,R) framework, time may, therefore,  undergo
small scale 
random fluctuations in the neighbourhood of every instant ($\bar t<<1$).
Two instants $t_{\pm}$  joined by an inversion,  can therefore be in a
continual 
process of transport between each other by randomly flipping their signs. 
The possibility of random flipping also indicates that there is an
inherent uncertainty in 
the actual determination of an instant close to $t=1$. In case one claims
that he/she is 
in the moment $t_{-}$, then there is a 50\%  chance that he/she is
actually 
at the moment $t_{+}$. 
From now on, we denote by $t$ this stochastic behaviour of time, when the
ordinary 
Newtonian time is denoted by $\eta$.

3.  The ordinary nonrandom variable (time) $\eta$, that we are accustomed
to 
at moderate scales should be retrievable from the stochastic $t$ in the
mean 
$\eta = <t>$ (upto a rescaling).  Physically, it means that the small
scale 
fluctuations near every point of $t$ cancel each other in such a way to
yield  
the average {\em coarse-grained} time sense $\eta$ to our
experience. These 
fluctuations would, however, become important to determine the small scale 
structure of time and hence of a dynamical  system. By inversion, these 
would also have nontrivial influences on the long time behaviour of the 
system. Incidentally, we note that the formal definition of inversion is
still valid for 
$\eta$. For a nonzero mean $\bar t$, for instance, inversion between 
$\eta_{\pm}=1\pm <\bar t>$  could then be implemented at a well determined 
instant. For a zero mean $\bar t \,(<\bar t>=0$) ,  on the other hand, 
the inversion constraint collapses to $\eta=1$. Thus allowing for 
a random, zero mean $\bar t$ is equivalent to probing small scale
structure of a point, $t=1$, 
where by point we mean a `physical point' which is meaningful only in the
context of 
an accuracy limit (level of resolution), in contrast to an ideal
structureless point . 
As an example, let us consider a situation when the 
measurement of a duration (an interval) is allowed only upto the first
decimal accuracy. 
In that case, points separated by a distance less than .05 are
indistinguishable, and should 
be treated as an equivalence class. In this sense, the point $t=1$ is an
equivalent class, 
where the symbol `1' is only a convenient representative of the class.  In
the ideal case 
of infinite accuracy, the class `1' has only one element in the ordinary
(linear) Calculus. 
In the presence of inversions, it now follows that 
even an ideal point ( in the limit of infinite accuracy) has infinite
number of nontrivial members 
separated by infinitesimally small, zero mean random scales. Since $\bar
t(t)$ is a 
positive, monotonically increasing stochastic  variable, the nontrivial
equivalence class 
of an ideal point has, at the least,  the cardinality of continuum. In
fact, it can be 
higher, as is shown in Sec.4. (It will be evident that zero mean is not a
necessary 
restriction.)

4. It turns out that  the concepts of inversion and stochastic time are
deeply related to 
the (practical ) limitation on the exact measurability of a duration. Let,
for instance, the 
moments $t_{-},\,t_{+}$  belong to the equivalence class of the `physical
point' $t=1$. Then for 
any practical purpose, $t_{-}=t_{+}=1$.  That means, in turn, that any
measurement  over a 
finite period of time would fail to make any distinction between
$t_{-},\,t_{+}$ and 1.  The 
associated fluctuations between $t_{-},\,t_{+}$ which remain imperceptible
over a pretty  
long time can, however, act as potential seeds for self-similar evolutions
over scales 
$t_n=\nu^n t$, which can grow slowly to interfere with each other, thereby
deflecting 
the standard nonrandom evolution, to a universal pattern of stochastic
fluctuations (c.f.,Sec.3).

5. It is worth comparing the present definition of a local time inversion
with the usual 
(global) time reversal (inversion) symmetry 
of an equation of the form eq(1). The usual time reversal symmetry means
that 
the system $x(t)$ evolves not only forward in time from $t$ to $t+h,h>0$, 
but it can also evolve backward; i.e., the state $x(t)$ can be
reconstructed from 
the state $x(t+h)$. The parameter $t$ in eq(1) is thus `non-directed', 
giving rise to the problem of time asymmetry. As remarked already, the new 
class of solutions indicates an inherent irreversibility in the time
sense.

\section{Fractal Solution}
\par  In the framework of SL(2,R) stochastic time, eq~(\ref{simo}),  
rewritten as 

\begin{equation}\label{stoch}
{{\rm d}x=(1+\kappa t)x\,{\rm d}t}
\end{equation}

\noindent assumes the status of a stochastic differential equation.
We now present a treatment of the above equation when the 
variable $t$ is assumed to act as the ordinary 'coarse grained' time
$\eta$, 
except in the vicinity of $t=1$ (the dominant scale of evolution). Thus 
over the period (0,1), the system evolves along the standard evolutionary 
path $x_0\equiv x_s\approx e^\eta$ ($\kappa$ being small). The small 
scale fluctuations in the system variable $x$, as inherited from the
fluctuations 
in $t$ would remain un-noticeable in comparison with this mean
evolution. One 
can, however, probe these fluctuations, provided the mean 
evolution $x_0$ is removed from the actual (total) system $x$ 
via the ansatz $x=x _0x_1$, where  $x_1$ is the purely `fluctuating' 
component satisfying the reduced equation
 
\begin{equation}\label{redu}
{{\rm d}x_1=\kappa t\,x_1{\rm d}t}
\end{equation}

\noindent To prove the self similarity of the small scale fluctuations,
and to see 
how these  grow and influence the late time properties of the mean
evolution,   
we probe the neighbourhood of $t=1$ more and more closely, using 
scales $t_n$,  separating in each step the 
relevant residual evolution into `mean' and `fluctuation' respectively. 
To see this explicitly, let us assume, for definiteness, that 
the actual removal of the mean in eq(\ref{stoch}) is accomplished at an 
instant $t_{-}\lessapprox 1$. We note ,however, that the specification 
of the exact moment is physically impossible. Thus the moment $t_{-}$ 
must have an inherent uncertainty, which we represent by an 
ansatz $t_{-}=1-r\bar\eta$, where, $r$ is a discrete variate with moments
$<r^n>
\approx r_0^n, \,<r>=r_0$, so that $<t_{-}>=\eta_{-}$. 
For the sake of clarity, one may also 
assume that $t_{+}$ and $ t_{-}$ belong to the physical equivalence 
class of 1(c.f., remark 4), when a point is defined correct to $m$ 
decimals (so that the accuracy limit is $10^{-m}$).
Because of the random inversions $t_{+}\leftrightarrow t_{-}$ 
in the time coordinate, the residual system pair $x_1(t_{-})$ and 
$x_1(t_{+})$ also undergo fluctuations between them.  However,  
these fluctuations would remain unobservable at least over a 
sufficiently long period 
of the form $0<\eta<1+a$, provided $r_0$ is sufficiently small and 
$a\sim 1/(10^m r_0)$. {\em Now utilising the 
golden mean partition of unity}: 
$\nu^2 +\nu=1,\,\nu>0$,  one can realise a scale changing
SL(2,R)  transformation

\begin{equation}\label{gold}
t_{+}=1+r\bar \eta\approx {\frac{1+\tilde r\nu\,\bar \eta_1}{1-\tilde
r\bar \eta_1}}
\end{equation}

\noindent where $\bar \eta_1=r_0\nu \bar \eta=\nu(\eta_{+}-1)<<1,\, \tilde
r=r/r_0$. 
Using  eq(\ref{gold}) and the inversion constraint ${\rm d}t_{-}=-{\rm
d}t_{+} 
=-r{\rm d}\bar \eta$, eq(\ref{redu})  gets transformed to

\begin{equation}\label{ssim}
-{{\rm d}x_1=\tilde r\lambda(1+\tilde r\nu T_1)\,x_1{\rm d}T_1}
\end{equation}

\noindent  where $T_1=\ln{\eta_1},\,\eta_1=1+\bar\eta_1= 1+\nu(
\eta_{+}-1)$ 
and $\lambda=\kappa/\nu$, when we make use of $\ln(1+\tilde
r\bar\eta_1)\approx 
\tilde r\bar\eta_1\approx \tilde r T_1$. This equation, 
valid close to $\bar \eta=0, (\eta=1)$, and self similar to
eq(\ref{stoch}) , describes 
the small scale evolution of the first generation fluctuation $x_1$. (One
can, indeed, recast 
eq(\ref{ssim}) exactly to the form eq(\ref{stoch}) in the stochastic time
variable 
$t_1=\tilde r T_1$.) The (-) sign  is a signature of inversion. Note that
this also avoids the 
possibility of a backward flow of time at the expense of deflecting the
direction of 
system evolution. As time $\eta$ continues 
to flow from $\eta_{+}\approx 1$ onwards,  dragging $\bar \eta_1$ along
with, the small 
scale fluctuation $x_1$  also gets amplified following eq(\ref{ssim}) 
and assumes the status of the original system 
variable $x$ over time $\eta\sim O(e(1+\nu)-\nu)$, when a 
second generation transition to the scale $T_2$ becomes permissible. 
Note that the self similarity of eq(\ref{ssim}), relative to the time
variable 
$t_1$, with eq(\ref{simo}) tells that $T_1$ would act as the ordinary time
for 
$T_1\in (0,1)$ for the evolution of $x_1$. Factoring $x_1=e^{-\lambda
t_1}x_2$, 
one thus gets the 2nd generation replica of eq(\ref{simo}) for $x_2$ in
the time 
variable $T_2$ and hence this method of self 
replication over scales $T_n$ could continue ad infinitum. An infinite
string of iterations, as 
above, thus leads to a new solution of eq(\ref{simo}) in the form 
\begin{equation}\label{fsol}
 x \propto e^{\eta-\tilde r\lambda( T_1- T_2 +\ldots)}
 \end{equation}
 
\noindent Note that $ T_2=\ln \eta_2, \,\eta_2=1+\nu( T_1-1)=1+\nu(
\ln(1+\nu (\eta-1))-1))$ 
and hence the nth generation scale $T_n$ is related to $\eta$ by  $n$ 
nested natural logarithms. Consequently, the over all fluctuations in the
system  
$x_f\propto
x/x_0=({\frac{\eta_2\eta_4\ldots}{\eta_1\eta_3\ldots}})^{\tilde
r\lambda}$, 
incorporating influences of all scales,  has the asymptotic form 
$x_f\sim ({\frac{\ln{\eta }\ln\ln\ln{\eta} \ldots}{\eta \ln
\ln{\eta}\ldots}})^{\tilde r\lambda}
\equiv \eta^{-\tilde r\mu}$, as $ \eta\rightarrow \infty$. 
Here, the exponent $\mu=\lambda(1-{\frac{\ln\sigma}{\ln\eta}}),
\, \sigma= {\frac{\ln{\eta }\ln\ln\ln{\eta} \ldots}{\ln
\ln{\eta}\ldots}}$, 
is a slowly varying function of $\eta$.
A few remarks are in order.

1. The choice of the moment $t=1$, around which the evolution is probed, 
 is for the sake of convenience. For any point $t_0\in (0,1)$, the
analysis 
proceeds with the rescaling $t\rightarrow t/t_0,\, \kappa\rightarrow
t_0^2\kappa$ 
in eq(\ref{redu}).

2. Because of the nested logarithms, contributions from higher order
scales 
are felt slower and slower. As stated above the first sign of fluctuation
is surfaced 
only if the system is allowed to evolve over a period $\eta_f\sim
1+10^{-m} r_0^{-1}$. 
As an example, in a `universe' where only the first order accuracy
($m=1$) is allowed, the 
fluctuation is first noticed around $\eta_f\sim 2$ for a $r_0\sim
0.1$. More  
generally, for a $r_0=10^{-(m+s)}$, 
$\eta_f$ could be arbitrarily large, for a large value of $s$, even in the
limit $m\rightarrow 
\infty$ of infinite accuracy. 
 
3. It is natural to interprete the new solution eq(\ref{fsol}), with self
similar fluctuations 
over the mean solution $x_0$, as a (random) fractal solution of the
equation 

\begin{equation}\label{simp}
{{\rm d}x\over {\rm d}t}=x 
\end{equation}

\noindent under time inversions. This is, however,  in contrast to our 
aim which was to find such a solution for eq(\ref{simo}), so that
stochastic fluctuations 
would have been around the standard solution $x_s$. In the following, we
resolve 
this dichotomy in a more general framework of fractal time, where the
parameter 
$\lambda$ is identified with an `apriori' scale factor of time, rather
than a system 
variable. We also show how the correct fluctuation pattern is obtained
when 
$\lambda$ acts as a true system variable.

\section{Fractal time}

\par We have shown how a stochastic behaviour could be injected to 
time via inversion, while time inversion is interpreted as a consequence 
of the practical limitation of exact measurability of a duration. 
( Stated more precisely, the possibility of time inversion raises the 
practical measurement limitation to the level of a theoretical 
principle.) The stochastic nature of time already endows time 
with fractal~-like characteristics. To investigate fractal properties 
in more details, let us begin by writing an {\em ansatz} for the
`physical' fractal time 
$t$ as an implicit random function of the ordinary time
$\eta: t_{\pm}(\eta)
=\eta(1\pm\kappa\eta t_{\mp}(\eta^{-1})) $. 
Here, $\kappa>0$ stands for an arbitrary, small 
but random, parameter. The nontrivial second factor would be responsible 
in determining the small scale structure of  the physical time in the
neighbourhood 
of every point $\eta$. The choice of sign $\mp$ in $t$ in the factor
indicates explicitly 
the possibility of inversion near $\eta=1$ (see below). Note also that
$t_{\pm}$ mimics 
the notation of Sec.2-3, thereby splitting every point of $\eta$ axis into
an 
equivalence class $\{t_{\pm}\}$ of infinite number of finely separated
points. 
Since each member of the class is a function of $\eta$, it has, at least,
the cardinality 
of continuum (c.f., remark 3, Sec.2). We show below that the actual
cardinality 
is $2^c$.

\par As it turns out, the ansatz represents {\em a new class of  exact, 
stochastic solutions} to the equation ${dx\over d\eta}=1$.  
To verify this, we note (suppressing the distinction temporarily)
that, by symmetry, both $t/\eta$ and 
$\eta\tilde t,\,\tilde t=t(\eta^{-1})$ satisfy coupled equations
of the form $\alpha=1+\kappa \beta$ and 
$\beta=1+\kappa \alpha$, 
hence $t(\eta)/\eta=\eta\tilde t(\eta)$ for all $t$.
Noting that ${{\rm d}\tilde t\over {\rm d\eta}}
=-\eta^{-2}{{\rm d}\tilde t\over{\rm d}\eta^{-1}}$, we get 
${{\rm d}t\over {\rm d}\eta}= (1+\kappa \eta\tilde t~)+
\eta\kappa \tilde t -\kappa {{\rm d}\tilde t\over {\rm d}\eta^{-1}}$, 
so that ,$({{\rm d}t\over {\rm d}\eta}-{t\over \eta})+
\kappa ({{\rm d}\tilde t\over {\rm d}\eta^{-1}}-\eta\tilde t~)=0$. It 
thus follows, $\kappa>0 $ being arbitrary,  that 

\begin{equation}\label{sode}
{\eta{{\rm d}t\over {\rm d}\eta}=t}
\end{equation} 

\noindent which is nothing but the desired equation in logarithmic
variables.
Note that for a nonrandom real parameter $\kappa$, 
one retrieves the standard solution $t=(1-\kappa)^{-1}\eta$. For a random 
$\kappa$, which arises naturally in the context of the present `local' 
definition of inversion, we, however, get a new {\em random
(fractal) solution}, 
which matches (approximately) with the standard linear solution only in 
the mean. We 
emphasise that the new solution is an {\em exact} solution of
eq(\ref{sode}). 
Because of its inherent scale-free nature, the solution must possess 
nontrivial fractal characteristics. Indeed, eq(\ref{sode}) tells that $\ln
t$, and hence 
$t/\eta$, must be a function of $\ln \eta$, so that
$t=\eta\phi(\ln\eta$). Here,  
$\phi(\ln\eta)=c(1+ \kappa \phi(\ln \eta^{-1}))$, and 
represents a nontrivial (random fractal ) solution of $\frac{{\rm
d}x}{{\rm d}\eta}=0$, 
$c$ being a real constant. 

\par Let us note that a straightforward iteration of the ansatz in the
form 
$t/\eta=1+\kappa+\kappa^2+\ldots$ would be, in general,  misleading 
because this geometric series in  $\kappa$ apparently hides the 
slow time dependence that is always present in any finite approximants 
of this infinite series. We have already shown how such a  slow, residual 
 time dependence in the $n$th approximant 
$S_n=1+\kappa+\kappa^2+\ldots+\kappa^n \eta\tilde t$
can influence the dynamics over time $\eta\sim <\kappa>^{-n}$, 
because of local time inversions. 

\par  {\em To explore the role of golden mean $\nu$}, {\em ab-initio} in
the present context, 
let us now reintroduce signs '$\pm$' to distinguish the variables
$t_{\pm}$. 
Let  $t_{+}/\eta=1+r\eta_1t_{-}(\eta_1^{-1})$ and
$t_{-}/\eta=1-r\eta_1t_{+}(\eta_1^{-1})$, 
where $\eta_1=k\eta,\,\kappa=rk, k$ is an ordinary constant, and $r
\,(\sim 1)$ 
is a random variable, analogous to one in Sec.3. Note that the present
form is slightly 
general from above, but, nevertheless, solves eq(\ref{sode}), because of
its scale free 
nature. This scale free property now tells that the limit of
$t_{+}(\eta)/\eta\equiv \phi(\eta)$ 
as $\eta\rightarrow \eta_0, 0< \eta_0< \infty $ is independent of
$\eta_0$. As a 
consequence, $\phi(\eta)$ is a {\em universal} random function, defined in
the 
vicinity of $\eta=1$, so that $t_{+}=\eta\phi(\eta),\, \phi(\eta)=
1+r\phi(\eta_1^{-1})$. Note that for a sufficiently small $k$, the
variation 
of $\phi$, which always remains of the order O(1),  is very small. 
Note that $t_{+1}(\eta)=kt_{+}(\eta)=t_{+}(\eta_1)$, by definition.
\par Let us now recall that  $\eta_1= 1 $ is an `ideal'  point in the
ordinary 
time axis, in contrast 
to the `physical  point' $t=1$, an equivalence class of members of the
form  
$\{t_{\pm}\}$ in the geometrical axis of the physical time. 
Now, as $\eta$ approaches $\infty $ crossing $k^{-1}$,  the rescaled 
variable $\eta_1 $ crosses $\eta_1 =1$,  running over points  such as 
$\eta_{1-}=1-\sigma$ to $\eta_{1+}=1+\sigma$. 
An ordinary time inversion $\eta_{1-}=\eta_{1+}^{-1}$ (c.f., remark
3,Sec.2), 
now induces a random inversion 
in the physical time $t_{-}(\eta_{1-}^{-1})\rightarrow t_{-}(\eta_{1+}): 
rt_{-}(\eta_{1+})=\eta_{1+}^2/t_{+}(\eta_{1+})$, so that
$r\phi(\eta_1^{-1})= 1/\phi(\eta_1)$. 
Note that the exact moment of inversion is uncertain 
because of the inherent randomness in the physical time due to the
r.v. $r$.
Consequently, one obtains $t_{+}/\eta=1+\eta_1/t_{+}(\eta_1)$, with
$\eta_1=1+\sigma, 
\sigma\rightarrow 0$. It thus follows that
$t_{+}=\phi\eta,\,\phi(\eta)=1+\nu$, which is true 
not only for $\eta\rightarrow \infty $, but, in fact, for {\em any} $\eta$ 
because of the universality of $\phi$. 

\par {\em To continue further, let us note that the above result, although
proves the unique 
role of the golden mean number in the framework of fractal time, also
presents us with 
a riddle.} Apparently, one would like to conclude that this solution
reproduces the 
standard solution $t=c\eta$ of eq(\ref{sode}). However, the emergence of
this 
unique special value $\nu$ is unclear in the ordinary framework. Recall
that the variables 
$t_{\pm}$ are intrinsically random. Further, no where in the above
analysis we have 
taken mean values to eliminate the underlying randomness.  
The only reasonable conclusion would therefore 
be the following: {\em the golden mean number $\phi$ represents a
universal  
random fractal function which is responsible for small scale random
fluctuations in 
the physical time $t$}.\footnote{ From now on, we denote by $\phi$ this
stochastic 
time-like feature of the golden mean. By the symbol $\nu$ we, however,
continue to 
refer to the usual `non-random' number ${\frac{\sqrt{5}-1}{2}}$.} 
 
To explain the assertion in detail, we need to proceed in 
a number of steps, establishing a number of  key results.

\par Let us begin by noting that the ordinary solution of
eq(\ref{sode}) defines 
a 1-1 ( the identity) mapping $R\rightarrow R$ of the ordinary real
set. The new class 
of random scale free solutions now  defines an extension of the ordinary
real 
set to the `physical ' real set $P$, say. The extended solution space of
linear ordinary   
differential equations now consists of solutions which are  (i) infinitely
differentiable 
and (ii) those which are finitely differentiable, stochastic functions. We
shall  verify shortly  
that the new implicitly defined functions are first order differentiable, 
with discontinuous second derivatives, at the moments 
when $\eta$ changes by inversions. Because of this extension, it is
natural to expect 
that the physical set $P$  contains new members not available to $R$.  To
show 
that $P$ indeed has nontrivial numbers, let us first    
distinguish a physical number $K_p$ from an ordinary positive real 
number $K$, where $K_p=K+N,\, N$ being  the neighbourhood of 1 
consisting of random physical numbers (fluctuations). Thus, a 
physical number $K_p$, representing a nontrivial equivalence class, is 
a non-singleton subset of $R_{+}$, the set of nonzero positive reals. 
Further, by definition $K_p=K\phi(\eta)$. 
\par Let $P(\tilde R)$ denote the power set of $\tilde R=R_{+}\cup
(-1,0),\, \tilde P$, 
being the corresponding extension, and $N$ the set of discrete subsets in
$\tilde R$. 
Then the cardinality of $N$ equals, $c$,  the cardinality of
continuum. Let 
$g:N\rightarrow (-1,0)$ denote a natural 1-1 correspondence.
Now, there exists an injection $f_1: P(\tilde R)\rightarrow \tilde P$,   
defined by $f_1(A)=g(A)$, when $A\in N$; $f_1(A)=K$, when
eq(\ref{sode}) admits infinitely 
differentiable solutions in $A$; and $f_1(A)=K_p$, otherwise. Conversely, 
one also finds an injection $f_2:\tilde P\rightarrow P(\tilde R)$, 
defined by  $f_2(K)=\{K\}$, but $f_2(K_p)=\{K,B\}$, where $B=(0,1)$ is the
interval 
where $K_p$ is defined. Hence, by the Schroeder-Bernstien theorem [7], the 
cardinality of the physical set $P$ equals $2^c$. 
Note that the mapping $g$ renders all possible discrete subsets of 
$P$ unphysical, transferring them to (-1,0). Further,
eq(\ref{sode}) admits infinitely 
differentiable solutions, over a given, {\em apriori} scale denoted as
$\eta$, 
provided the possibility of random scale dependence via inversions is
neglected. 
Consequently, the set $R$ is realized in $P$ in an approximate sense.
Moreover,  there exist  an uncountably more  nontrivial elements in the 
physical set in comparison to the real set. This means, in particular,
that there exist
physical numbers ( in the form of fluctuations) which are infinitely
smaller than any 
nonzero ordinary real numbers. By inversion, one then concludes that there
exist infinitely 
large physical numbers greater than any real number. Clearly,  all the set
theoretic results, 
valid in $R$ , get carried over to the physical set. In particular, the
physical set is 
partially ordered, when $k_p\leq K_p$ means $k\leq K$. 

\par Now to construct a nontrivial infinitely small fluctuation, let us
recall that a 
physical number is associated with an accuracy limit. Let
$k_p=rk(\eta^{-1})\eta$ be 
determined with an accuracy $10^{-m}$. Let sup$ [1/k]\approx 10^{(m+s)}$. 
Here, $[.]$ denotes the greatest integer function, and the supremum 
is defined on a bounded interval of $\eta$. Then, as 
$m\rightarrow \infty,\, s $ large, but finite, $k_p=0$, in the ordinary
sense, for 
finite $\eta$. But in the sense of a `physical limit' $\eta_1=10^{-m}\eta
\rightarrow 1,\, m\rightarrow \infty ,\,\eta\rightarrow \infty $, one gets
an arbitrarily 
small random number, by allowing $s$ to assume larger and larger values, 
but never exactly allowing $s=\infty $. Because of randomness, $k_p$ 
is not an ordinary real number. The physical limit tells that the ideal 
condition of infinite accuracy is realised only in an infinitely distant
time. In 
other words, any measurement process over a finite period of time can 
achieve only a finite degree of accuracy. We note that the physical set
$P$ 
has a structure analogous to the nonstandard real number set [5]. A more 
detail investigation of the relationship between the two approaches 
will be considered separately.

\par Next {\em we show how such an infinitely small fluctuation $k_p$ can
have 
nontrivial influence at the level of the ordinary scale, thereby 
re-deriving, in an alternative way, the small scale time evolution 
in the golden mean $\nu$.} 
Let us rewrite the physical time $t$ in the form 
$t=\eta \phi(\eta),\, \phi(\eta)=1+k_p\phi(\eta_1^{-1}), 
\, \eta_1=k\eta$. Since $t$ must be an exact solution of eq(\ref{sode}), 
one obtains $k_p{\frac{{\rm d}\phi}{{\rm d}\ln\eta_1}}=0$. It thus follows
that in the 
ordinary scale of $\eta: {\frac{{\rm d}\phi}{{\rm d}\ln\eta}}=0$, and
hence $\phi$ 
is an ordinary constant (having no evolution). However, this conclusion 
may not necessarily be true at the level of the  smaller scale $\eta_1$. 
In fact, as $\eta_1$ grows to order O(1), ${\frac{{\rm d}\phi}{{\rm
d}\ln\eta_1}}$, need 
not vanish, since $k_p$, though nonzero, becomes vanishingly small 
provided ${\frac{{\rm d}\phi}{{\rm d}\ln\eta_1}}\sim$ O(1). To verify
this, note that 
$\phi(\eta) -1=\bar\eta_1(\phi(\eta_1^{-1}))\,(\phi<1$ as long as
$\eta_1<1$),
$\bar\eta_1\approx 0$, so that 
in the limit of $\eta_1=1+\bar\eta_1\rightarrow 1$, 

\begin{equation}\label{nu}
{\frac{{\rm d}\phi(\eta_1^{-1})}{{\rm d}\ln\eta_1}}=\phi(\eta_1^{-1})
\end{equation}

\noindent which is of order O(1), and one retrieves a self similar 
replica of eq(\ref{sode}), over the scale $\eta_1$. Note that, as 
$\eta_1\rightarrow 1^-,\, \phi$ behaves as a small scale linear time 
so that $\phi-1\approx -{\rm d}\phi$. But as $\eta_{1-}\rightarrow
\eta_{1+}$, 
by inversion, (-) sign cancels so as to reproduce eq(\ref{nu}). It is now 
easy to verify the continuity of the first derivative at $\eta_1=1$, since 
$-{\frac{{\rm d}\phi(\eta_1^{-1})}{{\rm d}\ln\eta_1^{-1}}}={\frac{{\rm
d}\phi(\eta_1)}
{{\rm d}\ln\eta_1}}=\phi$. {\em The continuity of the second derivative
can not, 
however, be maintained because of the sign difference injected by
inversion.}

\par Note that the above analysis leads us to the fact, by yet another
route, that 
the golden mean $\nu=\phi-1$, must have small scale {\em intrinsic} time 
evolution. To state more clearly how a time sense gets attached to $\nu$, 
let us note that the intrinsic time must flow, in the neighbourhood of
$\eta(=1)$, 
in the form $\phi(\eta)=1+k_p\phi(\eta_1^{-1})$, till a random inversion 
near $\eta=k^{-1}$ carries it to the smaller scale $\eta_1$, to assume the
form 
$\phi(\eta)= 1+k(\phi(\eta_1))^{-1}$. As a consequence, a (linear) time
sense, self similar to 
the scale $\eta$ is generated over the scale $\eta_1$ , which persists  
upto order $\eta_1\sim k^{-1}$, preparing it for yet another replication 
on the second generation scale $\eta_2$ and so on, leading to an infinite
continued 
fraction $\phi_k =1+[k;k;k;\ldots]$ representation of the intrinsic time
flow, as the zeroth
generation ordinary time $\eta\rightarrow \infty $, exploring longer and
longer 
scales. Letting  $k=1$,  one gets {\em the golden mean flow of time}. Note
that the 
intrinsic sense of time is derived from the infinitely slow random
unfolding 
of cascaded scales hidden in the form of an infinite continued fraction. 
Among all these time-like 
continued fractions $\phi_k$, the golden mean continued fraction is
distinguished by 
its slowest possible convergence (unfolding) rate. We now show that
$\phi_k -1
\rightarrow \phi$ as $\eta\rightarrow \infty $. Note that after $n$th
inversion, $\eta$ 
reaches the scale $\eta_n$. Let $n_{\infty }$ denote the smallest
infinitely large 
natural number, exceeding all real numbers, in the physical set $P$, 
so that $\eta_{\infty }=k^{n_\infty }\eta$. One exhausts all possible
ordinary 
scales by utilising all the available $n_{\infty }$ number of  inversions
so 
that  the final replication leads to $\phi(\eta_{\infty
})=1+{\frac{1}{\phi(\eta_{\infty })}}$ , 
yielding $\nu$. It thus follows that although, in general, time may
flow following the steps of $\phi_k$, this scale dependent flow can in
fact continue 
at most upto a finitely many scales (in the context of physical
time). {\em The intrinsic 
flow, exploring smaller and smaller scales at slower and slower rates, 
as the ordinary zeroth generation scale $\eta\rightarrow \infty $,
would finally converge to the slowest possible golden mean flow.} Note
that once {\em the 
golden mean flow} is reached, no further scale replication is allowed,
since 
all the higher order scales would be physically indistinguishable
($\eta_{\infty +1}=
\eta_{\infty }$), from the stand point of the ordinary scale $\eta \sim
1$. However, because 
of self similarity, any scale $\eta_n$ could be considered as the zeroth
generation 
scale, the process of exploring the golden mean flow of time would remain 
unaltered.

\par We close this section with a few more remarks, highlighting a number
of important 
features of the golden mean.

1.  The golden mean time sense is intrinsic, since 
it is independent of an {\em apriori} time. Note that
$\phi(\phi(.))=\phi(.)$, where 
(.) indicates that $\phi$ could be a function of any $t_n$. In fact, the 
equation means  $\phi=\phi(\phi)$, thus eliminating the Newtonian external 
time $\eta$. (Recall that to avoid any confusion, we choose $\phi$ to
indicate the slow 
scale dependent (logarithmic) variation in the golden mean, while
$\nu^{-1}$ 
denotes the usual constant value of it, over a well defined given scale.)
However, an approximate, `coarse-grained' Newtonian sense of time is 
realised over a scale, when  the small scale fluctuations, because of 
local inversions, are ignored. In the following, we denote by $\phi(\phi)$
the 
set (equivalence class) of all possible intrinsic variables of the form 
$\{\phi(t_n)\}$. 

2. So far, we have not spelled out the form of the random variable $r$. A 
probability distribution satisfying the constraint $<r^n>\approx r_0^n,\,
<r>=r_0$ can be written as follows. Let the sample space of $r$ be 
$\{r_0^{s+1}\},\,s=0,1,2,\ldots$. The corresponding discrete probability 
function is defined by $p(r)=e^{-r_0^m}{\frac{r_0^{ms}}{s!}}, \,m$ being a  
positive integer. The distribution 
is Poisson-like, but not exactly the same, because of the special sample 
space, generating {\em the infinite set of scales}. Let $r_0\lesssim 1$,
so 
that $r^m_0<<1$, for a suitably large $m$. 
Thus $<r^n>=r_0^np(r_0), p(r_0)=e^{r_0^m(r_0^n-1)}
\approx 1$. Further, any number $r_0$ can be written as $r_0=\nu^a,\,
a=\log_{\nu}r_0$. Thus all the free parameters 
in the definition of the fractal time $t$ is determined self consistently 
in terms of $\nu$. The parameter $k=1$, since the scale factor is included 
in the sample space. One thus fixes $\lambda$ in eq(1) and (6) as 
$\lambda=r_0$. 

3. The golden mean $\nu$ remains constant over a scale  $\eta$, but,
nevertheless, 
enjoys intrinsic randomness. Because of the equality $\nu=(1-\nu)\Sigma
\nu^{2n}$, 
$\nu$ is realized as the expectation value of a r.v. $r$ with sample space
$\{\nu^n\}$
of scales, the corresponding probabilities being
$\{p(r)\}=\{(1-\nu)\nu^n\}$. 
Note that $\nu$ here denotes the exact value represented as $\frac{\sqrt
5-1}{2}$.
Let us consider the 
sample space $\{S_{m+n}\}$ of yet another r.v. $r_\nu$, for a sufficiently
large $m$, 
where $S_n$ is the $n$th approximants of the above series. The
corresponding 
probability distribution is $p(r_\nu)$. Consequently, $r_\nu$ realizes
higher 
precision values of $\nu$ with lower and lower probabilities. Clearly,
$\nu\equiv 
r_\nu$. This proof, being purely of number theoretic origin, tells that
measurements 
in a physical universe are inherently uncertain, since any measurable
quantity 
is a multiple of $\nu$. Note that the evaluation of the exact $\nu$ needs
a 
measurement process defined over an infinite amount of time, thus making
it 
physically impossible. The choice of the probability distributions may,
however, 
appear arbitrary. In Sec.6.2, we show how these discrete distributions
relate 
naturally to a universal density function for the class of fluctuations
considered 
in this work.

\section{Applications}

\par  In the following applications of the fractal time, we show how the 
formalism yields (i) a natural resolution to the generic $1/f$ signal
problem [1], 
and (ii)  the universal  probability density observed in the Bramwell-
Holdsworth- 
Pinton (BHP) fluctuations [2]. 

\par To begin with let us first reconsider eq(\ref{simo}) in the light of
fractal time $t$.  
Recalling  $t=\eta(1+ r\phi(\eta^{-1})),\, {\rm d}t=(1+ r\phi){\rm
d}\eta$, since 
${\frac{{\rm d}\phi}{{\rm d}\eta}}=0$. 
The equation (\ref{simp}) in the `physical' time $t$ is re-expressed in
the 
Newtonian time as 

\begin{equation}\label{simp1}
{{\rm d}x\over {\rm d}\eta}=(1+r\phi(\eta^{-1}))x 
\end{equation}

\noindent An exact solution of this equation, eq(\ref{fsol}), as an
implicit function 
of $\eta$  has already been obtained. An analogous form of the solution 
is now written as $x=e^{(\eta+\alpha(\eta))}\, \alpha(\eta)
=r\int\eta{\rm d}\phi(\eta^{-1})$, when eq(\ref{nu}) is used. Note that 
the infinite series of scales $T_n$, with alternating signs, provides an 
explicit representation of the intrinsic time variable $\phi$, when the
system 
is allowed to evolve over all the available scales in an infinite period
of the 
ordinary time $\eta$. The exponent $\alpha$ thus leads to the exponent
$\mu$ 
of eq(\ref{fsol}), when the integration is performed successively over
scales 
between two consecutive moments of inversion. It thus follows  that 
the method of  Sec.3 actually yields the fractal solutions to
eq~(\ref{simp}), when 
$\kappa$ is identified with one of the scales $\nu^n$. Moreover, a
solution of 
 eq(\ref{simo}) when $\kappa$ is treated as a system variable can be 
obtained as $x=e^{\eta+{\frac{1}{2}}\kappa\eta^2-\beta(\eta)}$, where 
$\beta(\eta)=r\int(\eta+2k\eta^2){\rm d}\phi(\eta^{-1}))$, neglecting
O($r^2$) 
corrections. Note that both the solutions indicate almost identical late 
time power law fluctuations. 

\subsection{1/f spectrum} 

\par To calculate the spectrum of the stochastic fluctuations, present
universally 
in linear equations of the form  
eq(\ref{simo}), we need to estimate the late time asymptotic form of 
the corresponding two-point correlation 
function $C(\eta)=<x(0)x_f(\eta)>=<x_f(\eta)>$, since $x(0)=1$. Assuming
that 
$|\lambda|\approx \nu^n$ (say), it follows that $C(\eta)\sim \eta^{-\mu}$, 
and hence the spectrum has the form $S(f)\sim 1/f^{1\mp\mu}$, $\mu$  
being the slowly varying function of Sec.3. Clearly this should be 
the generic form of the spectrum for a general  dynamical system of the
form 

\begin{equation}\label{gen}
{{\rm d}x\over {\rm d}t}=h(t)x 
\end{equation}

\noindent where  the time dependence in $h$ may have nonlinear influences: 
$h(t)=h(t, x)$. However, the nature of explicit nonlinearity in a system
is 
expected to get reflected in the exponent $\mu$. The generic logarithmic 
correction in $\mu$ provides significant  insights into the late time
features 
of the dynamics, which might get revealed in a time series over a number
of 
different scales. One example is treated below.

\subsection{Universal probability function}

Recently, a universal pattern of self similar fluctuations have been 
reported to occur in many Natural processes. Subsequently, 
a generic probability density function (PDF) of the form 

\begin{equation}\label{prob}
P(t)=Ke^{at-ae^t}
\end{equation}

\noindent is shown to be respected by the underlying dynamics [2], of 
apparently unrelated systems. Here, $t$ 
is a relevant fluctuating variable. Let us note that the solution
eq(\ref{fsol}) denotes 
the universal fluctuation 
pattern, at least, for those Natural processes which satisfy an equation
of the form 
eq(\ref{gen}). We now show how the above PDF is naturally realized for
this universal 
fluctuation.

\par Let us begin by noting that the infinite alternating series of scale
dependent terms 
in eq(\ref{fsol}) gives the complete fluctuation spectrum $x_f$  of a
linear system, 
which is self similar over all these scales and is revealed over an
infinite period of 
the ordinary time $\eta$. However, the first two terms in the series are
sufficient 
to capture the {\em generic} features of the fluctuation $x_f$ , because
the scale 
generating r.v. $r$ (c.f., remark 2, Sec.4) induces a higher order
(stochastic) scale 
dependence on each of the scales $T_n$, thereby inscribing a complete
replica, of the 
total fluctuation $x_f$ over the scale $T_1$ (say), even in a finite
period of $\eta$. 
Further, the (-) sign between two consecutive scales $T_1$ and $T_2$ is a
nontrivial 
signature of inversion. Thus it suffices for us to consider a renormalised
fluctuation of the form 
$\tilde x_f\propto e^{-r(T_1-T_2)}$ ($\lambda=r_0$), for a finite
$\eta$. Accordingly, 
the generic PDF corresponding to $\tilde x_f$ should be identical with the
same for $x_f$. 

\par Let, for definiteness, that the random scales $r=\{\nu^{n-1}\}$ be
distributed with 
probabilities $p(r)$ introduced in remark 3, Sec.4. Because of the
logarithmic scale 
dependence $T_2=\ln(1+\nu(T_1-1))$, one gets $\tilde x_f\propto
e^{r(T_2-(1+\nu)e^{T_2})}$. 
Dropping the higher order scale dependence introduced by the $\nu$
dependent factor, we 
get finally  $\tilde x_f\propto e^{r(T_2-e^{T_2})}$.
Now in a practical situation, the fluctuation $\tilde x_f$ is represented
in the form 
of a time series record over a period of time. The probability that a
randomly sampled 
observation $\tilde x_{fn}$ is drawn from the scale $r_n$, in the sample
space  of $r$, 
now equals the product of the probability of selecting the scale $r_n$ and
the 
conditional probability that the observation actually comes from the said
scale, given 
that the scale has been chosen already. But the conditional probability is
nothing 
but the correlation function $C(T_2)=<\tilde x_{fn}(T_2) \tilde
x_f(0) >=<\tilde x_{fn}(T_2)>,\, 
 \tilde x_f(0)=1,\,\tilde x_{fn}\propto e^{r_n(T_2-e^{T_2})}$. Thus the
probability of 
drawing a random sample from the scale $r_n$ equals 
$P(r_n)\propto p(r_n)e^{r_n(T_2-e^{T_2})}$. Hence the grand universal
probability  
that the time series reveals the whole spectrum of fluctuations over all
possible 
scales is obtained as   

\begin{equation}\label{pro}
P_u\propto \Sigma_0^{\infty} P(r_n)=
(1-\nu)\Sigma_0^{\infty} \nu^ne^{(1+\nu)\nu^n(T_2-e^{T_2})}
\end{equation}

\noindent Clearly, apart from a multiplicative factor, which could be 
fixed from the normalisation condition,  the zeroth order term of this
infinite series 
representation of the universal PDF agrees well with that in ref.[2], 
which was obtained from an  
approximate argument (the factor $a=\pi/2$ in the exponential gets
replaced here 
by $1+\nu$). The higher order terms in the infinite series represent
corrections, 
which are likely to give more accurate fits of the time series records
for Natural 
processes, as noted in ref.[2].

\par Now to explain the reason of the matching, we note that the factor
$1+\nu$ in eq(\ref{pro}) 
actually realises $T_2$ as $T_1(\equiv (1+\nu)T_2)$ and $T_1$ as
$\eta$. In the case of 
an explicitly nonlinear system given by  
eq(\ref{gen}) with nonlinearity coupling $k\sim 1$, the system experiences
fluctuations 
at a time $\eta\sim 1$. The scale $T_1$ then corresponds to the 2nd
generation fluctuation 
in a corresponding time series record. It thus follows that the zeroth
order PDF would 
have the form $e^{T_1-e^{T_1}}$. Let us note that the moments of the model 
fluctuating variate in ref.[2] possesses the generic property
$<r^n>\propto r_0^n$. However, 
the corresponding PDF is obtained from a quantum field theoretic
consideration of a critical 
magnetic model which is based purely in the Newtonian time frame. Now, the
relationship 
between an `ordinary' dynamical variable $Q_0$, following an evolutionary
equation of the form 
eq(\ref{gen}), but in the ordinary time, and the corresponding `physical'
variable $Q_p$ is  
given by $Q_0\propto Q_p^\nu$, since in the logarithmic scale $\ln
t=(1+\nu)\ln\eta$. An application 
of this conversion rule thus leads to 

\begin{equation}\label{pro1}
P_{BHP}\propto e^{(1+\nu)(T_1-e^{T_1})}
\end{equation}
\noindent which completes the derivation of the BHP probability function.

\par To proceed further, let us now re-derive the generic PDF from an
alternative method. 
This will reveal a subtle relationship between the universal PDF and the
set of discrete 
distributions we have chosen as examples.  
Let $\eta_n=\nu^n\eta$, where $\eta\in (0,\infty )$ be nonrandom, and
$\nu$ denote an 
approximate value of the golden mean. The scales $\eta_n$ are random,
because of the 
randomness in $\nu$ and is assumed to follow the distribution of remark 2,
Sec.4 ($m=1$), 
so that $\eta_1$ is realized with probability $P_1=e^{-\nu}\nu$. Since
$\nu$ is approximate, 
it will now undergo evolution in the physical set following, for instance,
the scale free  
representation $\nu_p=\nu\phi(\eta_1^{-1})
=\nu/(1+\phi(\eta_{11}^{-1}))=\nu/(1+/\{1+\phi(\eta_{12}^{-1})\})$,
etc. We may assume that 
the linear sense of time generated by $\phi$ at the scale of $\eta$ is
denoted by $\eta$ 
itself. As $\eta$ flows from $\eta\approx 0$  slowly, the intrinsic
evolution in $\nu$ splits 
the scale $\eta_1$ into an infinite set of tiny scales
$\eta_{1n}=\nu^n\eta_1$, each of 
which will further undergo finer levels of subdivisions, and so on. 
The stochastic evolution of $\nu$ thus fractures the scale $\eta_1$ at the 
neighbourhood of every  point, thereby raising 
it to the level of a continuous r.v. with the PDF $P_1(\eta_1)\propto 
e^{-\nu\phi(\eta_1^{-1})}\phi(\eta_1^{-1})$, a Gamma distribution. We note 
that El Naschie [8] has already used a Gamma distribution to derive the 
Hausdorff dimension of the fluctuating (Cantorian) spacetime as $4+\nu^3$. 
He has also indicated how the mass spectrum of all known elementary 
particles could be determined using the golden mean $\nu$. 
(The uncertainty exponent of the fractal time is obtained independently by
the author 
as   $\nu$ [10].)  

\par All the higher order scales will similarly undergo infinitesimal 
fluctuations and hence finally be distributed following the above 
Gamma density function, since $\nu^n_p=\nu^n\phi(\eta_n^{-1})$. 
Note that all these $\phi(\eta_n^{-1})$ functions correspond to 
different scale dependent realizations of the same universal function 
$\phi(\phi)$ in the limit of infinite time. 
Letting $\nu\phi=e^{\tilde \phi}$, we reproduce the universal 
PDF for scale free (self similar) fluctuations, 
$P(\tilde \phi)\propto e^{\tilde \phi-e^{\tilde\phi}}$. 
But $\tilde\phi$ is again a realisation of $ \phi(\phi)$, by eq(\ref{nu}) 
and remark 1, Sec. 4.

\section{Concluding remarks}

The extension of the real set  to the physical set provides 
a dynamical representation of  the number system, each member of which is 
associated with an equivalence class of fluctuating elements separated by 
infinitesimally small scales. The fundamentally stochastic nature of the 
golden mean number renders accurate measurements in the physical set
impossible. 
Consequently, the physical universe based on the physical set would
consists of 
intrinsic changes and fluctuations. It is hard to imagine any
fundamentally constant 
physical quantity in this universe. The real number set is an incomplete 
realization of the physical set, when the possibility of infinitesimal
changes by 
local inversions are ignored. However, because of the new exact class of
solutions, 
$P\equiv R$. In the midst of all these changes and 
approximations, there exists, however,  one symbol of perfection in the
form of the 
golden mean equation $\phi(\phi)^2 +\phi(\phi)=1$, being engraved
fundamentally 
in the formalism of {\em  the SL(2,R) Calculus}. A more detailed, in depth
analysis of this 
Calculus,  will be presented in a subsequent paper, where the status of
well known 
theorems such as the Picard's existence and uniqueness theorem will be
examined. 
Let us only remark here that the present class of solutions are not in
contradiction with 
the Picard's theorem, the scope of which gets extended in the
SL(2,R) formalism. 
The present  formalism is likely to initiate a new approach in 
understanding the origin and dynamics of the nonlinear phenomena in
Nature.

\section*{Acknowledgements}
I am thankful to Prof. M. S. El Naschie for suggesting some improvements
in the 
original version of of the paper.

\end{document}